\documentclass[12pt]{article}
\usepackage{amssymb}
  \usepackage{amsthm,amsfonts}
  \usepackage{amsmath}
    \usepackage{latexsym}
\usepackage[applemac]{inputenc}
\newcommand{\bea}   {\begin{eqnarray}}
\newcommand{\eea}   {\end{eqnarray}}
\newcommand\ei{\end{itemize}}
\newcommand{\beq}{\begin{equation}}
\newcommand{\eeq}{\end{equation}}

\newcommand{\beano}{\begin{eqnarray*}}
\newcommand{\enano}{\end{eqnarray*}}

\newcommand{\bee}{\begin{enumerate}}
\newcommand{\ene}{\end{enumerate}}

\newcommand{\bei}{\begin{itemize}}
\newcommand{\eni}{\end{itemize}}

\newcommand{\bra}{\begin{array}}
\newcommand{\era}{\end{array}}
\newcommand{\bqn}{\begin{eqnarray}}
\newcommand{\eqn}{\end{eqnarray}}
\newcommand\ben{\begin{enumerate}}
\newcommand\een{\end{enumerate}}

\newcommand{\DEF}{\stackrel{\mathrm{def}}{=}}
\newcommand{\deq}{\stackrel{\mathrm{def}}{=}}

\def\NU{\Upsilon}

\def\BC{\mathbb C}
\def\_\BC{\mathbbi C}

\def\SD{\rtimes}

\def\R{\mathbb{R}}
\def\N{\mathbb{N}}
\def\C{\mathbb{C}}

\def\ds {de Sitter}

\newcommand{\be}{\beta}
\newcommand{\ga}{\gamma}

\newcommand{\vth}{\vartheta}
\newcommand{\al}{\alpha}
\newcommand{\de}{\delta}

\newcommand{\lga}{\longrightarrow}

\begin{document}
\renewcommand{\thefootnote}{\fnsymbol{footnote}}

\thispagestyle{empty}

\title{A natural  fuzzyness of de Sitter  space-time}
\author{Jean-Pierre Gazeau\thanks{{\em e-mail: gazeau@apc.univ-paris7.fr}}
 ~and Francesco
Toppan\thanks{{\em e-mail: toppan@cbpf.br}}
\\ \\
{\it $~^\ast$Laboratoire Astroparticules et Cosmologie (APC, UMR 7164),}\\ {\it Boite 7020 Universit\'e
Paris Diderot Paris 7,}\\{\it P10, rue Alice Domon et L\'eonie Duquet 75205, Paris Cedex 13, France.}\\
\\
{\it $~^{\dagger}$TEO, CBPF, Rua Dr.} {\it Xavier Sigaud 150,} \\{\it cep 22290-180, Rio de Janeiro (RJ), Brazil.}}
\maketitle
\begin{abstract}
A non-commutative structure for de Sitter spacetime is naturally introduced by replacing (``fuzzyfication") the classical variables of the bulk in terms of the dS analogs of the Pauli-Lubanski operators. The dimensionality of the fuzzy variables is determined by a Compton length and
the commutative limit is recovered for distances much larger than the Compton distance. The choice of the Compton length determines different scenarios. In scenario I the Compton length is determined by the limiting Minkowski spacetime. A fuzzy dS in scenario I implies a lower bound (of the order of the Hubble mass) for the observed masses of all massive particles (including massive neutrinos) of spin $s>0$. In scenario II the Compton length is fixed in the de Sitter spacetime itself and grossly determines the number of finite elements (``pixels" or ``granularity") of a de Sitter spacetime of a given curvature.
\end{abstract}
\vfill

\rightline{CBPF-NF-007/09}

\newpage

\section{Introduction}

Within the framework of Quantum Physics in Minkowskian space-time, an elementary particle, say a quark, a lepton, or a  gauge boson, is identified through some basic attributes like mass, spin, charge and flavour. The (rest) mass is certainly the most basic attribute for an elementary particle. Now, for a  particle of non-zero mass, its relation to space-time geometry on the quantum scale  is irremediably limited by  its (reduced) Compton wavelength
\begin{equation}
\label{compmass}
\lambda_{\mathrm{cmp}} = \frac{\hbar}{mc}.
\end{equation}
It is sometimes claimed that $\lambda_{\mathrm{cmp}}$  represents ``the quantum response of mass to local geometry'' since  it is considered as the cutoff below which quantum field theory, which can describe particle creation and annihilation, becomes important.

Now we know, essentially since Wigner, that mass and spin attributes of an ``elementary system'' emerge from space-time symmetry. These arguments rest upon the  Wigner classification of the Poincar\'e unitary irreducible representations (UIR) \cite{newtonwigner,wigner}:
the UIR's of the Poincar\'e group are completely characterized by
 the eigenvalues of its two Casimir operators,
 the quadratic  $\mathcal{C}^{0}_2 = P^{\mu}\,P_{\mu} = {P^0}^2 - \mathbf{P}^2$ (Klein-Gordon operator) with eigenvalues $
\langle \mathcal{C}^{0}_2 \rangle =  m^2\, c^2$ and the
quartic $\mathcal{C}^{0}_4 = W^{\mu}
W_{\mu}, \ W_{\mu} = \frac{1}{2}\epsilon_{\mu \nu \rho \sigma}
J^{\nu \rho}P^{\sigma}$ (Pauli-Lubanski operator)
with eigenvalues (in the non-zero mass case)
$\langle \mathcal{C}^{0}_4  \rangle = -m^2\, c^2\, s(s+1)\hbar^2$.

These results lead us to think that the mathematical structure to be retained in the description of mass and spin is the  symmetry group, here the Poincar\'e group $\mathcal{P}$, of space-time and not the space-time itself. The latter may be described  as the coset $\mathcal{P}/L$, where $L$ is the Lorentz subgroup. On the other hand we know that a UIR of $\mathcal{P}$ is the quantum version (``quantization'') of a co-adjoint orbit \cite{kirillov} of $\mathcal{P}$, viewed as the classical phase space of the elementary system. The latter is also  described as a coset: for an elementary system with non-zero mass and spin  the coset is $\mathcal{P}/(\mbox{time-translations} \times SO(2))$.  This coset is by far more fundamental than space-time.

Since a co-adjoint orbit may be viewed as a phase space or set of initial conditions for the  motion of an elementary particle, and so is proper to the latter, the existence of a ``minimal'' length  provided by  its Compton wavelength leads us to consider
the space-time as a ``fuzzy manifold" proper to this system. This raises the  question to establish a consistent  model of  a fuzzy Minkowski space-time issued from the Poincar\'e UIR associated to that elementary system.  The answer is not known in the case of  a flat geometry. However note that the Pauli-Lubanski vector components $W^{\mu}$ could be of some use in the ``fuzzyfication'' of the light-cone in Minkowski, just through the replacement $x^{\mu} \rightarrow  W^{\mu}$ and by dealing with massless UIR's of the Poincar\'e group in such a way that the second Casimir $\mathcal{C}^{0}_4$ is fixed to zero. The non-commutativity stems from the rules $\left\lbrack W^{\mu}, W^{\nu} \right\rbrack = - i  \epsilon_{\mu \nu  \rho \sigma} P^{\rho}W^{\sigma}$ and the covariance is granted thanks to the rules
$\left\lbrack W_{\mu}, P_{\nu} \right\rbrack = 0$ and $\left\lbrack J_{\mu \nu }, W_{\rho} \right\rbrack = i\left(\eta_{\mu \rho} W_{\nu} - \eta_{\nu \rho} W_{\mu}\right)$.

In this note we show that there exists a consistent way for defining such a structure for any ``massive system'' if we deal instead with a de Sitter space-time.

The organization of the paper is as follows. In Section {\bf 2} we recall the basic features of the de Sitter space-time and of its application to the cosmological data suggesting an accelerating universe. In Section {\bf 3} we compactly present the main properties of the de Sitter group UIR's. In section {\bf 4} we discuss the contraction limits of the de Sitter UIR's to the Poincar\'e UIR's. The main results are discussed in Section {\bf 5} and {\bf 6}. In Section {\bf 5} a non-commutative structure is naturally introduced in dS spacetime by assuming the bulk variables being replaced by ``fuzzy" variables (similarly to the analogous non-commutative structure of the fuzzy spheres) which, in a given limit, recover the commutative case. The ``de Sitter fuzzy Ansatz" implies a lower bound (of the order of the observed Hubble mass $1\approx 1.2 ~10^{-42} GeV$) for the observed masses
of the massive particles of spin $s>0$. In Section {\bf 6} the ``de Sitter fuzzy Ansatz" is applied to the desitterian physics and its cosmological applications. In the Conclusions we discuss the implication of these results and outline possible developments.

 \section{The de Sitter hypothesis}
 
In a curved background the mass of a test particle can always be  considered as the rest mass of the particle as it should  locally hold  in a tangent minkowskian space-time.
However, when  we deal with a de Sitter or Anti de Sitter background, which are constant curvature space-times, another way to examine  this concept of mass is possible and should also be considered.  It is precisely based on symmetry considerations in the above Wigner sense, i.e. based on the existence of the simple de Sitter or Anti de Sitter groups that are both one-parameter deformations of the Poincar\'e group.
 We recall that the de Sitter [resp. Anti de Sitter]  space-times are  the  unique
maximally symmetric solutions of the vacuum Einstein's equations with positive [resp.
negative] cosmological constant $\Lambda$.  Their
respective invariance (in the relativity or kinematical sense) groups
are the ten-parameter \ds ~$SO_{0}(1,4)$ and Anti \ds ~$SO_{0}(2,3)$ groups.
Both may be seen as   deformations of the
proper orthochronous Poincar\'e group $\R^{1,3}\rtimes\, SO_{
0}(1,3)$, the kinematical group of Minkowski.
 Exactly like for the flat case, dS and AdS space-times can be identified as cosets $SO_{ 0}(1,4)/\mbox{Lorentz}$ and $SO_{ 0}(2,3)/\mbox{Lorentz}$ respectively, and coadjoint ($\cong$ adjoint) orbits of the type $SO_{ 0}(1,4)/SO(1,1)\times SO(2)$ (resp. $SO_{0}(2,3)/SO(2)\times SO(2)$) can be viewed as phase space for ``massive" elementary systems with spin in dS (resp. AdS).
  Since the beginning of
the eighties the de Sitter space has been
considered as a key model in inflationary cosmological scenario
where it is assumed that the cosmic dynamics was
dominated by a term acting like a cosmological constant.
More recently, observations on far high redshift
supernovae, on galaxy clusters and on
cosmic microwave background radiation (see for instance \cite{LAMBDA}), suggest an
accelerating universe. This can be explained in a satisfactory way  with such a
term. This constant, denoted by $\Lambda$,  is linked
to the (constant) Ricci  curvature $4 \Lambda$ of these
space-times and it allows to introduce the fundamental curvature
or inverse length $ H c^{-1}=\sqrt{\vert \Lambda\vert/3} \equiv R^{-1}$, ($H$ is
the Hubble constant).

To a  given  (rest minkowskian) mass $m$ and to  the existence of  a  non-zero curvature   is naturally associated the typical dimensionless parameter for dS/AdS perturbation of the minkowskian background:
\begin{equation}
\label{param}
\vth_m \deq  \frac{\hbar \sqrt{\vert \Lambda} \vert}{\sqrt{3}mc}=  \frac{\hbar\, H}{ mc^2} = \frac{m_H}{m}\, ,
\end{equation}
where we have  also  introduced a ``Hubble mass" $m_H$ through
\begin{equation}
\label{hubmass}
m_H = \frac{\hbar H}{c^2}.
\end{equation}

We can also introduce the Planck units, defining the regime in which quantum gravity becomes important, which are determined in terms of $\hbar$, $c$ and the gravitational constant $G$, through the positions
\bea
length &:& l_{Pl} = \sqrt{\frac{\hbar G}{c^3}}\approx 1.6\times 10^{-33} cm,\nonumber\\
mass &:& m_{Pl} = \sqrt{\frac{\hbar c}{G}},\approx 2.2 \times 10^{-5} g \approx 1.2 \times 10^{19} GeV/c^2,\nonumber\\
time &:& t_{Pl}= \sqrt{\frac{\hbar G}{c^5}}\approx 5.4 \times 10^{-44} s,\nonumber\\
temperature &:& T_{Pl} = \sqrt{\frac{\hbar c^5}{G{k_B}^2}}\approx 1.4 \times 10^{32} K.
\eea
The observed value of the Hubble constant is
\bea
H \equiv H_0&=& 2.5 \times 10^{-18} s^{-1}.
\eea
Associated to the Planck mass, we have  the dimensionless parameter $\vartheta_{Pl}$ through
\bea
\vartheta_{Pl}&=& \frac{m_{H_0}}{m_{Pl}}= t_{Pl}H_0\approx 1.3 \times 10^{-61},
\eea
while
\bea
\Lambda t_{Pl}^2 c^2&=& \Lambda {l_{Pl}}^2 = {9}{\vartheta_{Pl}}^2 \approx 1.6 \times 10^{-121}
\eea
(namely, the cosmological constant is of the order $10^{-120}$ when measured in Planck units) and
\bea
\frac{R}{l_{Pl}} &=& (H_0t_{pl})^{-1}\approx 0.8\times 10^{61}.
\eea
As a consequence, if $l_{Pl}$ is a minimal discretized length, $(\frac{R}{l_{Pl}})^3\approx 10^{180}$ measures the number
of discrete elements (``atoms") in a quantum de Sitter universe.\par

We give in Table below  the values assumed by the quantity $\vth_m$ when $m$ is taken as  some known masses and $\Lambda$ (or $H_0$) is given its present day estimated value. We  easily  understand from this table that the currently estimated value of the cosmological constant has no practical effect on our familiar massive fermion or boson fields. Contrariwise, adopting  the  de Sitter point of view appears as inescapable when we deal with  infinitely small masses, as is done in standard inflation scenario.

\begin{table}[ht]
\begin{center}
{ \begin{tabular}[c]{|c|c|}\hline
Mass $m$ & $\vth_m \approx $\\ \hline
Hubble mass $m_{\Lambda}/\sqrt{3}\approx 0.293\times 10^{-68}$kg & 1 \\ \hline
up. lim. photon mass $m_{\gamma}$& $0.29 \times 10^{-16}$\\ \hline
up. lim. neutrino mass $m_{\nu}$& $0.165 \times 10^{-32}$\\ \hline
electron mass $m_e$& $0.3 \times 10^{-37}$\\ \hline
proton mass $m_p$& $0.17  \times 10^{-41}$ \\ \hline
$W^{\pm}$  boson mass & $0.2  \times 10^{-43}$\\ \hline
Planck mass $M_{Pl}$& $0.135  \times 10^{-60}$\\ \hline
\end{tabular}}
\label{masstable}
\caption{Estimated values of the dimensionless physical quantity $ \vth_m \deq  \frac{\hbar \sqrt{\vert \Lambda} \vert}{\sqrt{3}mc}=  \frac{\hbar\, H_0}{ mc^2} \approx 0.293\times 10^{-68}\times m_{\mathrm{kg}}^{-1} $ for some known masses $m$ and the present day estimated value of the Hubble length $ c/H_0 \approx 1.2\times 10^{26} \mathrm{m}$ \cite{pada}.}
\end{center}
\end{table}

\section{1+3 De Sitter geometry, kinematics and dS UIR's}

Geometrically, de Sitter space-time  can be described as
an one-sheeted hyperboloid $\mathcal{M}_H$ embedded in a five-dimensional Minkowski
space (the bulk):
 $$\mathcal{M}_H \equiv \{x \in \R^5 ;~x^2 \DEF x\cdot x=\eta_{\alpha\beta}~ x^\alpha
x^\beta =-H^{-2} \equiv -R^2\}$$
$$\alpha,\beta=0,1,2,3,4, \
 \eta_{\alpha\beta}=diag(1,-1,-1,-1,-1)\, , \quad x := (x^0, \vec{x}, x^4).$$

 A global causal ordering exists on
the de Sitter manifold: it  is induced from that one in the ambient
spacetime ${\R}^{5}$ : given two events $x,y\in M_{H}$,
 $$x\geq y \  \mathrm{iff} \   x-y\in\overline {V^+}$$ where $$\overline
{V^{+}}=\{x\in{\R}^{5}: x\cdot x \geq 0,\ {\rm sgn}\, x^{0}={+}\}$$
is the future cone in ${\R}^{5}$. One says that
$\{y\in M_H: y\geq x\}$ (resp. $\{y\in M_H:
y\leq x\}$) $\equiv $ closed causal future (resp. past) cone of  point $x$ in
$X$.  Two events $x,y\in M_H$ are  in ``acausal relation"
or ``spacelike separated" if they belong to the intersection of the
complements of above sets, {\it i.e. } if $(x-y)^2 = -2(H^{-2} +
x\cdot y) < 0$.

There are ten Killing vectors generating  a Lie algebra isomorphic to $so(1,4)$:
\beq
\nonumber K_{\alpha \beta} = x_{\alpha}\partial_{\beta} -
x_{\beta}\partial_{\alpha}\, . \eeq
At this point, we should be aware that there is  no globally time-like Killing
vector in de Sitter,  ``time-like'' (resp. ``space-like'')
referring to the Lorentzian   four-dimensional
 metric induced by that of the bulk.
  The de Sitter group is $G=SO_{ 0}(1,4)$ or its universal covering, denoted $Sp(2,2)$,
 needed for dealing with half-integer spins.

 Quantization (geometrical or  coherent state or something else) of de Sitter classical phase spaces leads to their
quantum counterparts, namely the quantum elementary systems associated in a biunivocal way to
the UIR's of the de Sitter group $SO_{ 0}(1,4)$ or $Sp(2,2)$. The ten
Killing vectors  are represented as (essentially) self-adjoint operators in  Hilbert
space of (spinor-)tensor valued functions on $M_H$, square integrable with respect to some
invariant inner (Klein-Gordon type) product :
$$
K_{\al \be} \rightarrow L_{\al \be} = M_{\al \be} + S_{\al \be}\, ,
$$
where  $M_{\al \be}=-i(x_{\al}\partial_{\be} - x_{\be}\partial_{\al})$ (orbital part), and
  $S_{\al \be}$ (spinorial part) acts on indices of functions in a certain permutational way. Note the usual
  commutation rules,
  \begin{equation}
  \label{lalbecom}
\left\lbrack L_{\al \be} ,L_{\ga \de} \right\rbrack = -i\left(\eta_{\al \ga}L_{\be \de} - \eta_{\al \de}L_{\be \ga}- \eta_{\be \ga}L_{\al \de}+ \eta_{\be \de}L_{\al \ga}\right) .
\end{equation}
 Two Casimir operators exist whose eigenvalues   determine  the UIR's :
\begin{equation}
\label{fixcas1}
\mathcal{C}_2 = - \frac{1}{2} L_{\al \be}L^{\al \be}\, ,
\end{equation}
\begin{equation}
\label{fixcas2}
\mathcal{C}_4 = - W_{\al} W^{\al}, \ W_{\al} = -
\frac{1}{8}\epsilon_{\al \be \ga \delta \eta} L^{\be \ga}L^{\delta \eta}\, ,
\end{equation}
where $(W_{\al})$ is the dS counterpart of the Pauli-Lubanski operator.
These $W_{\al}$'s transform like vectors:
  \begin{equation}
  \label{extra1}
\left\lbrack L_{\al \be} ,W_{\ga}\right\rbrack = i\left(\eta_{\be \ga}W_{\al} - \eta_{\al \ga}W_{\be}\right)\, .
\end{equation}
 In particular we have
\begin{equation}
W_a = i\left\lbrack W_{0}, L_{a 0} ,\right\rbrack\, , \quad a= 1,2,3,4\, .
\end{equation}
The $W_{\al}$'s commute as:
\begin{align}
 \label{extra2}
\nonumber \left\lbrack W_{\al} ,W_{\be}\right\rbrack &= - i\epsilon_{\al \be \al_1 \al_2 \al_3}W^{\al_1}L^{\al_2\al_3}\\
&= -i\left(L^{\ga \de}L_{\ga \de} - 3\right)L_{\al \be} + \frac{i}{2}\left\lbrace L^{\ga \de}, \left\lbrace L_{\al \ga}, L_{\be \de}\right\rbrace\right\rbrace\, .
\end{align}
The algebra defined by the $L_{\alpha\beta}$ and the $W_\alpha$ generators (regarded as primitive generators) is a non-linear finite
$W$-algebra \cite{BOERTJIN}.
This algebra respects the grading $[L_{\alpha\beta}]=2$, $[W_\alpha]=3$, $[[,]]=-1$.\par
As it is the case with non-linear $W$-algebras, (\ref{lalbecom}), (\ref{extra1}) and (\ref{extra2}) can be linearized (the ``unfolded" version) at the price of
introducing an infinite number of generators regarded as primitive generators. E.g., the r.h.s. of Equation (\ref{extra2}) can be
written as $-i\epsilon_{\alpha\beta\gamma_1\gamma_2\gamma_3}Z^{\gamma_1\gamma_2\gamma_3}$, where $Z^{\gamma_1\gamma_2\gamma_3}$ can be identified
with $W^{\gamma_1}L^{\gamma_2\gamma_3}$ ($[Z^{\gamma_1\gamma_2\gamma_3}] = 5$). An infinite tower of extra primitive generators
have to be introduced to close the algebra linearly.\par
It is convenient to reexpress the $L_{\alpha\beta}$, $W_\alpha$ generators of the finite $W$-algebra in their $SO(4)$ decomposition,
($a,b=1,2,3,4$), given by $T_a$, $L_{ab}$, $Z$, $W_a$, where
\begin{equation*}
T_\alpha = L_{0\alpha}\, , \quad
Z= W_0\, .
\end{equation*}
Due to the fact that the $W_a$'s arise from the commutator $[T_a,Z]$, we can regard the finite non-linear $W$-algebra with $T_a$, $L_{ab}$, $Z$, $W_a$
primitive generators as an unfolded version of the finite non-linear $W$-algebra with $T_a$, $L_{ab}$, $Z$ primitive generators.

The operator  $W_0$ is the difference of two commuting $su(2)$-Casimir. To get this, we start from the expression:
  \begin{equation*}
W_0 = -(L_{12}L_{34}+ L_{23}L_{14} + L_{31}L_{24}) = - \mathbf{ J}\cdot \mathbf{ A}\, .
\end{equation*}
The operators  $\mathbf{ J} = (L_{23}, L_{31}, L_{12})^T$ and  $\mathbf{ A} = (L_{14}, L_{24}, L_{34})^T$
represent  a basis for the maximal compact  subalgebra $\mathfrak{k} \cong so(4)$:
\begin{equation*}
[J_i,J_j] = i J_k\, , \ [J_i, A_j]= i A_k\, , \ [A_i,A_j] = iJ_k\, ,
\end{equation*}
with $(i,j,k)$ even permutation of $(1,2,3)$.
Introducing the two commuting sets of $su(2)$ generators:
  \begin{equation*}
 \mathbf{N}^{\mbox{\tiny L}}:= \frac{1}{2}(\mathbf{A} + \mathbf{J}), \, \mathbf{N}^{\mbox{\tiny R}}:= \frac{1}{2}(\mathbf{A} - \mathbf{J})\, , \ \left\lbrack N_i^{\substack{\mbox{\tiny L}\\
\mbox{\tiny R}} } , N_j^{\substack{\mbox{\tiny L}\\
\mbox{\tiny R}} }\right\rbrack =\pm iN_k^{\substack{\mbox{\tiny L}\\
\mbox{\tiny R}} }\, ,
\end{equation*}
we obtain
  \begin{equation*}
W_0 = - \mathbf{J}\cdot \mathbf{A} =   - \mathbf{A}\cdot \mathbf{J} = (\mathbf{N}^{\mbox{\tiny L}})^2 -  (\mathbf{N}^{\mbox{\tiny R}})^2 .
\end{equation*}
In consequence, as an operator on a direct sum of $SU(2)$ UIR's its spectrum is  made of the numbers $j_l(j_l + 1) - j_r(j_r + 1)$, $j_l, j_r \in \N/2$.
A complete classification \cite{DIX}  of the  de Sitter UIR's is precisely based on the following property. Let $Sp(2,2) \ni g \mapsto \rho(g) \in \mathrm{Aut}(\mathcal{H})$ a UIR of $Sp(2,2)$ acting in a Hilbert space $\mathcal{H}$.
Then the restriction to the maximal compact subgroup $K \deq  SU(2) \times SU(2)$ is completely reducible:
\begin{equation*}
\mathcal{H} = \oplus_{(j_l,j_r)\in \Gamma_{\rho}}\mathcal{H}_{j_l,j_r}\, , \quad \mathcal{H}_{j_l,j_r}\cong\C^{2j_l+ 1}\times \C^{2j_r + 1}\, ,
\end{equation*}
where $\Gamma_{\rho} \subset \N/2$ is the set of pairs $(j_l,j_r)$ such that the UIR $D^{j_l} \otimes D^{j_r}$ of $ K  $ appears once and only once in the the reduction of the restriction $\rho\mid_{K}$.
 Let $p = \inf_{(j_l,j_r) \in \Gamma_{\rho}}(j_l +j_r)$ and $q_0 = \min_{\substack{
(j_l,j_r) \in \Gamma_{\rho}\\ j_l +j_r = p}}(j_r-j_l)$,  $q_1= \max_{\substack{
(j_l,j_r) \in \Gamma_{\rho}\\j_l +j_r = p}}(j_r-j_l)$.
Then we have the following exhaustive  possibilities:
\begin{itemize}
  \item[(i)] $q_1 = p$ and $q_0 =-p$, which correspond to elements of the principal and complementary series, denoted by $\NU_{p, \sigma}$ where $\sigma \in (-2, + \infty)$ (with restrictions according to the values of $p$ in $\N/2$);
  \item[(ii)] $q_1 = p$ and $ 0< q_0 \equiv q \leq p$, which correspond to elements of the discrete series, denoted by $\Pi^{+}_{p,q}$;
  \item[(iii)] $q_0 = - p$ and $ 0< - q_1 \equiv q \leq p$, which correspond to elements of the discrete series, denoted by $\Pi^{-}_{p,q}$;
  \item[(iv)] $q_0 = q_1 = 0 \equiv q$ , which correspond to elements lying at the bottom of the discrete series, denoted by $\Pi_{p,0}$.
\end{itemize}
Let us now give more details  on these three different types of representations.

\subsubsection*{``Discrete series''  $\Pi^{\pm}_{p,q}$}

Parameter $q$ has a spin meaning and the two  Casimir are fixed as
  \begin{align*}
\label{}
\mathcal{C}_2 &= (-p(p+1) - (q+1)(q-2)) \mathbb{I}\, ,   \\
    \mathcal{C}_4&= (-p(p+1)q(q-1)) \mathbb{I}\, .
\end{align*}
We have to distinguish between
\bei
\item[(i)]{\it the scalar case}
$\Pi_{p,0}$, $p=1,2, \cdots$, which are not square integrable,
\item[(ii)] {\it the spinorial case} $\Pi^{\pm}_{p,q}$, $q>0$,
$p= \frac{1}{2}, 1, \frac{3}{2}, 2, \cdots$, $q=p, p-1, \cdots, 1$ or $\frac{1}{2}$. For $q = 1/2$ the representations
$\Pi^{\pm}_{p,\frac{1}{2}}$ are not square-integrable.
\eni

\subsubsection*{``Principal series'' $U_{s,\nu}$}
\begin{equation*}
\NU_{p=s, \sigma = \nu^2 + \frac{1}{4}}\equiv U_{s,\nu}\, , \  q = \frac{1}{2} \pm i \nu\, ,\ \sigma = q(1-q)\, .
\end{equation*}
$p=s$ has a spin meaning and  the two Casimir are fixed as
\begin{align*}
\mathcal{C}_2 &=  (\sigma + 2 -s(s+1) )\mathbb{I}= (9/4 + \nu^2 -s(s+1) )\mathbb{I}\, , \\
\mathcal{C}_4 &=  \sigma \, s(s+1) \mathbb{I} = (1/4+ \nu^2) \, s(s+1) \mathbb{I}\, .
\end{align*}
We have to distinguish between
\bei
\item[(i)] $\nu \in \R$ (i.e., $\sigma \geq 1/4$), $s=1,2, \cdots$, for the integer spin principal series,
\item[(ii)] $\nu \neq 0$ (i.e., $\sigma > 1/4$), $s= \frac{1}{2}, \frac{3}{2}, \frac{5}{2} \cdots$,
 for the half-integer spin principal series.
\eni
In both cases, $U_{s,\nu}$ and $U_{s,-\nu}$ are equivalent.
In the case $\nu = 0$, i.e. $q = \frac{1}{2}$,   $s= \frac{1}{2}, \frac{3}{2}, \frac{5}{2} \cdots$, the representations are direct sums of two UIR's
 in the discrete series:
 \begin{equation*}
U_{s,0} = \Pi^{+}_{s,\frac{1}{2}} \bigoplus \Pi^{-}_{s,\frac{1}{2}}\, .
\end{equation*}
\subsubsection*{``Complementary series'' $V_{s,  \nu}$}
$$\NU_{p=s, \sigma = \frac{1}{4} - \nu^2}\equiv V_{s,  \nu}, \, q = \frac{1}{2} \pm  \nu\, , $$
$p=s$ has a spin meaning and  the two Casimir are fixed as
\begin{align*}
\mathcal{C}_2 &= (\sigma + 2 -s(s+1)) \mathbb{I}= (9/4 - \nu^2 -s(s+1)) \mathbb{I}\, , \\
\mathcal{C}_4 &=\sigma \,s(s+1)\mathbb{I} = (1/4 - \nu^2) \,s(s+1)\mathbb{I}.
\end{align*}
We have to distinguish between
\bei
\item[(i)] {\it the scalar case} $V_{0,\nu}, \, \nu \in \R, \, 0 < \vert \nu \vert < \frac{3}{2}$ (i.e., $-2< \sigma < 1/4$),
\item[(ii)] {\it the spinorial case} $V_{s,\nu},  \, 0 < \vert \nu \vert < \frac{1}{2}$  (i.e., $0< \sigma < 1/4$), $s = 1, 2, 3, \dotsc$.
\eni
In both cases, $V_{s,\nu}$ and $V_{s,-\nu}$ are equivalent.

\section{Contraction limits or  desitterian Physics from the point of view of a Minkowskian observer}

On a geometrical level,  $\lim _{ H \to 0} \mathcal{M}_H = \mathcal{M}_0$, the Minkowski spacetime tangent
to $\mathcal{M}_H$ at, say, the de Sitter origin point $O_H$. Then, on an algebraic level, $\lim _{ H \to 0}Sp(2,2) = {\cal P}^{\uparrow}_{+} (1, 3) = \mathcal{M}_0  \SD SL(2,\BC)$, the Poincar\'e group.
The ten de Sitter Killing vectors contract  to their Poincar\'e counterparts $K_{\mu \nu}$, $\Pi_{\mu}$, $\mu =
0, 1, 2, 3$, after rescaling the four $K_{4\mu} \lga \Pi_{\mu} = H
K_{4\mu} $ (``space-time contraction'').  On a UIR level, the question is mathematically more delicate.
\subsubsection*{The massive case}
For  the massive case, the  principal series representations  only are
involved  (from which the name  ``de Sitter massive
representations''). Introducing  $\nu $ through $\sigma
= \nu^2 + 1/4$ and the Poincar\'e mass $m=\nu H$, we have the following result \cite{MINI,GAHURE}:
\beq
\nonumber
U_{s, \nu} \lga_{H\to 0, \nu \to \infty} {c_>\cal P}^{>}(m,s)
\oplus c_<{\cal P}^{<}(m,s),
\eeq
where one of the ``coefficients'' among $c_<, c_>$ can be fixed to 1 whilst the other one vanishes and where ${\cal P}^{\stackrel{>}{<}}(m,s)$ denotes
the positive (resp. negative) energy Wigner UIR's of the Poincar\'e
group with mass $m$ and spin $s$.

\subsubsection*{ The massless case}

 Here we must distinguish between
the  scalar massless case, which involves the unique complementary series
UIR $\NU_{0,0}$ to be contractively Poincar\'e significant,
and the helicity $=s$ case where are involved all representations
$\Pi^{\pm}_{s,s}, \ s>0$ lying at the lower limit of the discrete
series.
The arrows $\hookrightarrow $ below designate unique
extension. ${\cal P}^{\stackrel{>}{<}}(0,s)$ denotes the Poincar\'e massless case
with helicity $s$. Conformal
invariance leads us to deal also with the discrete series
representations
(and their lower limits) of the (universal covering of the)
conformal group or its double covering $SO_0(2,4)$ or its fourth
covering $SU(2,2)$ \cite{barbohm}. These UIR's are denoted  by
${\cal C}^{\stackrel{>}{<}}(E_0,j_1, j_2)$, where $(j_1,j_2) \in
\N/2 \times \N/2$ labels the UIR's of $SU(2) \times SU(2)$ and
$E_0$ stems for the positive (resp. negative) conformal energy.
\bei
\item[•] Scalar massless case :
\eni
{\footnotesize
\beq
\nonumber \left. \begin{array}{ccccccc}
& & {\cal C}^{>}(1,0,0)
& &{\cal C}^{>}(1,0,0) &\hookleftarrow &{\cal P}^{>}(0,0)\\
\NU_{0,0} &\hookrightarrow & \oplus
&\stackrel{H=0}{\longrightarrow} & \oplus & &\oplus \\
& & {\cal C}^{<}(-1,0,0)&
& {\cal C}^{<}(-1,0,0) &\hookleftarrow &{\cal P}^{<}(0,0),\\
\end{array} \right. \eeq}
\bei
\item[•] Spinorial massless case :
\eni
{\footnotesize
\beq
\nonumber \left. \begin{array}{ccccccc}
& & {\cal C}^{>}(s+1,s,0)
& &{\cal C}^{>}(s+1,s,0) &\hookleftarrow &{\cal P}^{>}(0,s)\\
\Pi^-_{s,s} &\hookrightarrow & \oplus
&\stackrel{H=0}{\longrightarrow} & \oplus & &\oplus \\
& & {\cal C}^{<}(-s-1,s,0)&
& {\cal C}^{<}(-s-1,s,0) &\hookleftarrow &{\cal P}^{<}(0,s),\\
\end{array} \right. \eeq
\beq
\nonumber \left. \begin{array}{ccccccc}
& & {\cal C}^{>}(s+1,0,s)
& &{\cal C}^{>}(s+1,0,s) &\hookleftarrow &{\cal P}^{>}(0,-s)\\
\Pi^+_{s,s} &\hookrightarrow & \oplus
&\stackrel{H=0}{\longrightarrow} & \oplus & &\oplus \\
& & {\cal C}^{<}(-s-1,0,s)&
& {\cal C}^{<}(-s-1,0,s) &\hookleftarrow &{\cal P}^{<}(0,-s),\\
\end{array} \right. \eeq
}

We can see from the above that there is energy ambiguity in de Sitter relativity,
exemplified by the possible breaking of dS irreducibility into a direct sum of
two Poincar\'e UIR's with positive and negative energy respectively.
This phenomenon is linked to the existence in the de Sitter group of
the discrete symmetry that
sends any point $(x^0, {\cal P}) \in M_H$  into its mirror image $(x^0, -{\cal P}) \in M_H$ with respect
to the $x^0$-axis.
Under such a symmetry the four generators
$L_{0a}$, $a = 1,2,3,4$, (and particularly $L_{04}$ which contracts
to energy operator!) transform into their respective opposite
$-L_{0a}$, whereas the six $L_{a b}$'s remain unchanged.
All representations that are not listed  in the above contraction limits   have either non-physical Poincar\'e contraction
limit or have no contraction limit at all.

In order to get a unifying description of the dS-Poincar\'e contraction relations,  the following  ``mass'' formula has been proposed by Garidi \cite{GAR1} in terms of the dS UIR parameters $p$ and $q$:
\begin{equation}
\label{garidimass}
m^2_{\mathrm{Gar}}= \langle \mathcal{C}_2 \rangle_{\mathrm{dS}} - \langle {\mathcal{C}_2}_{p=q} \rangle_{\mathrm{dS}}= [(p - q)(p + q - 1)] m_H^2\, , \quad m_H =\hbar H/c^2.
\end{equation}
 The minimal value assumed by  the eigenvalues of the first Casimir in the set of UIR in the discrete series is precisely reached at $p=q$, which corresponds to the ``conformal'' massless case.
The Garidi mass has the advantages to encompass all  mass formulas introduced within a de-sitterian context, often in a purely mimetic way in regard with their minkowskian counterparts.

Now, given a minkowskian mass $m$ and a ``universal'' length $R =\sqrt{3/ \Lambda}= c\,H^{-1}$, nothing prevents us to consider the dS UIR parameter $\nu$ (principal series), specific of a ``physics'' in constant-curvature space-time, as meromorphic functions of the dimensionless physical (in the minkowskian sense!) quantity $ \vth_m$, as was introduced in Equation (\ref{param})  in terms of  various other quantities introduced in this context,
 Note that $\vth_m$  is also the ratio of the Compton length of the  minkowskian object of mass $m$ considered at the limit with the universal  length $R$ yielded by dS geometry.
 Thus,  we may consider the following Laurent expansions of the dS UIR parameter $\nu$ (principal series) in a certain neighborhood of $\vth_m = 0$:
\begin{equation}
\label{laurentnu}
 \nu =     \nu (\vth_m)=   \frac{1}{\vth_m} + e_0 + e_1 \vth_m + \cdots e_n (\vth_m)^n + \cdots\, ,
\end{equation}
with $\vth_m \in (0, \vth_{\mathrm{max}})$ (convergence interval).
The  coefficients $e_n$ are pure numbers to be determined.
 We should be aware that nothing is changed in the  contraction formulas from the point  of view of a minkowskian observer, except that we allow to consider  positive as well as negative values of $\nu$ in a (positive) neighborhood of $\vth_m = 0$: multiply (\ref{laurentnu}) by $\vth_m$ and go to the limit $\vth_m \to 0$.
We recover asymptotically the relation
\begin{equation}
\label{dsnumass1}
m= \vert \nu \vert  \hbar H/c^2 = \vert  \nu \vert  \frac{\hbar}{c} \sqrt{\frac{ \vert \Lambda\vert}{3}}.
\end{equation}
As a matter of fact, the Garidi mass is a good example of such an expansion since it can be rewritten as the following expansion in the parameter $\vth_m \in ( 0,1/\vert s- 1/2 \vert]$:
\begin{equation}
   \nu =  \sqrt{\frac{1}{\vth_m^2} - (s-1/2)^2}
  = \frac{1}{\vth_m} -(s-1/2)^2\left(\frac{\vth_m}{2} + O(\vth_m^2)\right),
\end{equation}
Note the particular symmetric place occupied by the spin $1/2$ case with regard to the scalar case $s=0$ and the boson case $s=1$.

\section{Fuzzy de Sitter space-time with Compton wavelength}

Examining the equation (\ref{fixcas2}) that fixes the value of the quartic Casimir in terms of the operators $W_{\al}$ it is tempting, if no natural,  to introduce a non-commutative structure for the dS spacetime by replacing the classical variables of the bulk $x^\alpha$ with the suitably normalized $W^\alpha$ operators through the ``fuzzy" variables ${\widehat x}^\alpha$
\bea
x^\alpha &\rightarrow&{\widehat x}^\alpha= rW^\alpha,
\eea
where $r$ ($[r]=l$) has been introduced for dimensional reasons. In principle a different $r$ has to be introduced for any given irreducible representation characterized by $p,q$ (therefore $r\equiv r_{p,q}$).
\par
The non-commutativity reads as follows
\bea
[{\widehat x}^\alpha, {\widehat x}^\beta ]&=& -i r\epsilon_{\alpha\beta\gamma_1\gamma_2\gamma_3}{\widehat x}^{\gamma_1}L^{\gamma_2\gamma_3}
\eea
and goes to zero in the limit $r\rightarrow 0$.
\par
On the other hand, the dS inverse curvature $R$, arising from the classical constraint
\bea\label{curveq}
-x_\alpha x^\alpha &=& R^2,
\eea
is replaced in the ``fuzzy" case by the equation
\bea\label{quantum curveq}
-{\widehat x}_\alpha {\widehat x}^\alpha &=& -r^2p(p+1)q(q-1),
\eea
inducing the identification
\bea\label{curvatureandnc}
R&=& r\sqrt{-p(p+1)q(q-1)}.
\eea
What is $r$? A natural interpretation consists in assuming it to be a Compton length $l_{\mathrm{cmp}}$ of the associated particle.
The most immediate approach then consists in looking at the Minkowski limit fixing the Minkowskian observational mass $m_{\mathrm{obs}}$ as the limit
of the Garidi mass. In this case we have to work with the principal series which allows taking the limit $\nu\rightarrow \infty$.
We get that (\ref{garidimass}) can be rewritten as
\bea
m_{\mathrm{Gar}}&=& \frac{\hbar}{cR}\sqrt{(s-\frac{1}{2})^2+\nu^2}.
\eea
In agreement with (\ref{laurentnu}), we assume for $\nu$ a dependence on $R$ such as
\bea
\nu(R)&=& c_{-1}R+c_0+\frac{c_1}{R}+\frac{c_2}{R^2}+\ldots\, ,
\eea
and we can safely take the $R\rightarrow \infty$ limit in the r.h.s. and obtain
\bea
m_{\mathrm{obs}} &=& \frac{\hbar}{c}c_{-1}.
\eea
We are now in the position to use as $r$ in (\ref{curvatureandnc}) the Compton length associated to the observational
mass $m_{\mathrm{obs}}$. Since we are working with the principal series, (\ref{curvatureandnc}) can be in this case expressed as
\bea
R_0 &=& \frac{\hbar}{m_{\mathrm{obs}} c} \sqrt{s(s+1)(\frac{1}{4}+\nu^2)}\, ,
\eea
where $R_0 = c/H_0$.
The physical interpretation of the (observational) parameter $\nu$ is the fact that it connects the observational curvature
of the dS spacetime with the observational mass of the elementary system (for the given spin $s>0$).
We obtain
\bea
\nu^2 &=& \left(\frac{R_0^2m_{\mathrm{obs}}^2c^2}{\hbar^2}-\frac{s(s+1)}{4}\right)\frac{1}{s(s+1)}.
\eea
Since $\nu^2$ must be positive we end up with a constraint on the observational masses of the massive elementary systems with spin $s>0$.
The constraint is given by the relation
\bea\label{lowerbound}
m_{\mathrm{obs}} &\geq& \frac{\hbar}{2c}\sqrt{s(s+1)}\frac{1}{R_0} =  \frac{m_{H_0}}{2}\sqrt{s(s+1)}.
\eea
One can say that for all known massive particles this lowest bound is  of the order of the observed Hubble mass $\approx 1.6\, 10^{-42}$GeV.
Combining the upper limit of Table 1 and this lowest limit, we obtain the allowed mass range for massive neutrinos: $$\approx 1.2 \,  10^{-42}\,\mathrm{GeV} \leq m_{\mathrm n} \leq \approx 9.7 \, 10^{-11}\,\mathrm{GeV}\, .$$
\par
The lower bound (\ref{lowerbound}) for the observational masses of spin $s>0$ particles is based on the present-day value for the Hubble constant and the associated curvature of the universe. It admits, however, a reversed lecture. Starting from the known observational masses of the particles with positive spin, it could allow putting an upper bound (which can have cosmological implications) on the curvature of the dS universe.\par
It is worth pointing out that the ``de Sitter fuzzy Ansatz" implies a lower bound for the observed masses of all massive particles
of positive spin $s$. The lower bound does not depend on the electric charge (positive, negative or vanishing) of the particles. It depends smoothly
on their spin $s$. For large spin $s$ the lower bound grows linearly with $s$.\par
We refer to the construction of the fuzzy de Sitter spacetime discussed in this Section as ``Scenario I". In this scenario the Compton length has been defined in the limiting Minkowski spacetime. In the next Section we will discuss another scenario (referred to as ``Scenario II"), such that the Compton length is defined in the de Sitter spacetime itself.

 \section{Fuzzy de Sitter and Garidi mass}

Another interpretation of $r$ consists in assuming it to be a Compton length $l_C$ of the associated particle with its desitterian Garidi mass. The Compton length is in this case defined in the de Sitter spacetime itsef. We refer to this scenario as ``Scenario II".
For the irreducible representation characterized by $p,q$ the Compton length is
\bea
l_C &=& \frac{\hbar}{m_{\mathrm{Gar}}c}\, .
\eea
\par
Scenario II offers different possibilities of constraining the relevant $p,q$ UIR's which satisfy some given properties. We explicitly discuss two such cases. In the first case we have
\bea
l_C&=& \frac{\hbar}{cm_H}\sqrt{\frac{1}{(p-q)(p+q-1)}}= l_{Pl}\frac{1}{\vartheta_{Pl}} \sqrt{\frac{1}{(p-q)(p+q-1)}}
\eea
and
\bea\label{ratiocurvatureplanck}
\frac{R}{l_{Pl}}&=& \frac{1}{\vartheta_{Pl}} \sqrt{\frac{-p(p+1)q(q-1)}{(p-q)(p+q-1)}}.
\eea
For the principal series we can set
\bea
p&=& s,\nonumber\\
q&=& \frac{1}{2} +i\nu,
\eea
where $s$ is the spin and $\nu\in{\mathbb R}$. For this series we obtain that (\ref{ratiocurvatureplanck}) reads as
\bea\label{ratio}
\frac{R}{l_{Pl}}&=& \frac{1}{\vartheta_{Pl}} \sqrt{\frac{s(s+1)+\frac{s(s+1)}{4\nu^2}}{1+\frac{4s^2-4s+1}{4\nu^2}}}\approx
\frac{1}{\vartheta_{Pl}} \sqrt{s(s+1)} .
\eea
The last equality holds in the limit $\nu\rightarrow\infty$.\par
On the other hand, it is possible to choose $\nu\in{\mathbb R}$ such that the r.h.s. of (\ref{ratio}) does not depend on $s$.
For $s=1$ we get $\sqrt{\frac{s(s+1)+\frac{s(s+1)}{4\nu^2}}{1+\frac{4s^2-4s+1}{4\nu^2}}}=\sqrt{2}$ no matter which is the
value of $\nu$. In order to reproduce this value for any positive (half-)integer spin $s$, $\nu$ must be given by
\bea
\nu &=& \sqrt{\frac{7s^2-9s+2}{4s^2+4s-8}}.
\eea
The above equation always admits real solutions
(for $s=\frac{1}{2}$, $\nu=\sqrt{\frac{3}{28}}$, for $s=\frac{3}{2}$, $\nu=\sqrt{\frac{17}{28}}$, for $s=2$, $\nu=\frac{3}{4}$,
for $s=\frac{5}{2}$, $\nu=\sqrt{\frac{103}{28}}$, $\ldots$ with, in the limit $s\rightarrow \infty$, $\nu\rightarrow \sqrt{\frac{7}{4}}$).\par
In the massless case we have $m_{Gar}=0$, which implies either $p=q$ or $p=1-q$. In both cases
\bea
R &=& r\sqrt{p^2(1-p^2)},
\eea
which is positive only for $p=\frac{1}{2}$ (we are dealing with the ${\Pi^\pm}_{\frac{1}{2},\frac{1}{2}}$ irrep). Therefore
\bea
R&=& r \sqrt{\frac{3}{16}}.
\eea
In order to obtain the ``universal formula"
\bea\label{universal}
\frac{R}{l_{Pl}}&=& \frac{1}{\vartheta_{Pl}}\sqrt{2}
\eea
we have to fix $r$, the Compton length $l_C$ of the massless spin $\frac{1}{2}$ system, to be given by
\bea
r\equiv l_C &=& \frac{l_{Pl}}{\vartheta_{Pl}}\sqrt{\frac{32}{3}}.
\eea
Summarizing the results obtained so far, in the first case of Scenario II we can constrain the $p,q$ UIR's such that the ratio
$\frac{R}{l_{Pl}}$ between the de Sitter curvature and the Planck length becomes universal (i.e., the allowed $p,q$ produce
formula (\ref{universal})).\par

In a second case for Scenario II, the Compton length for the Hubble mass is associated with the curvature $R$ through
\bea\label{hubblecmp}
m_H &=& \frac{\hbar}{c R}.
\eea
By assuming (\ref{hubblecmp}) we can reexpress the Garidi mass $m_{\mathrm{Gar}}$ in terms of $p$, $q$ and the curvature $R$ (besides the constants $c$ and $\hbar$). We obtain
\bea \label{garidiandcurvature}
m_{\mathrm{Gar}} &=& \frac{\hbar}{cR}\sqrt{(p-q)(p+q-1)}.
\eea
Since
\bea
R&=& r \sqrt{-p(p+1)q(q-1)},
\eea
with $r$ the Compton length associated to the Garidi mass ($r=\frac{\hbar}{m_{\mathrm{Gar}}c}$), we end up with the equality
\bea
R&=& R\sqrt{\frac{-p(p+1)q(q-1)}{(p-q)(p+q-1)}}.
\eea
In this case the constraint
\bea
\frac{-p(p+1)q(q-1)}{(p-q)(p+q-1)}&=&1
\eea
has to be satisfied, giving a restriction on the allowed $(p,q)$ UIR's.\par
For the principal series ($p=s$, $q=\frac{1}{2}+i\nu$) the restriction corresponds to the equation
\bea
s(s+1)(\frac{1}{4}+\nu^2)&=&(s-\frac{1}{2})^2+\nu^2,
\eea
which implies
\bea
\nu^2 &=& \frac{1}{2}\sqrt{\frac{3s^2-5s+1}{s^2+s-1}}.
\eea
The above equation finds solutions for any positive spin $s>0$.
\par

\section{Conclusions}

In this work we described a natural fuzzyness of the de Sitter spacetime. In analogy to what happens in the case of the fuzzy spheres, a non-commutative structure for dS can be introduced by replacing the classical variables of the bulk with the dS analogs of the Pauli-Lubanski operators, suitably normalized by a dimensional parameter which plays the role of the Compton length
and measures the granularity of the non-commutative dS space-time.
 We adopt the point of view of the elementary system described by UIR's of the de Sitter group (parametrized by the pair $(p,q)$)
 and determine the mathematical and physical consequences of the ``de Sitter fuzzy Ansatz". Two different scenarios have been investigated according to the possibility that the Compton length of the elementary system is described in the
 Minkowski limit of the de Sitter spacetime (Scenario I; in this case the suitable UIR's of the de Sitter spacetime belong to the principal series and the UIR's of the Poincar\'e group are recovered in the limit $\nu\rightarrow\infty$) or in the de Sitter spacetime itself (Scenario II). The ``fuzzy dS Ansatz" in Scenario I has some observational implications. The observed masses of all massive particles with spin $s>0$ admits a lower bound of the order of the Hubble mass. For large $s$ the lower bound grows linearly with $s$. The lower bound is only function of the spin (not of the charge of the particles) and applies in particular to the case of the massive neutrinos. In scenario II the Compton length is fixed in the de Sitter spacetime itself and grossly determines the number of finite elements (``pixels" or ``granularity") of a de Sitter spacetime of a given curvature.
 In Scenario II the allowed UIR's (parametrized by $p$ and $q$) are determined in accordance with the properties that have to be required for the fuzzy dS spacetime. We discussed in particular two cases. A first case such that the ratio $\frac{R}{l_{Pl}}$ between the $dS$ curvature and the Planck length is universal. A second case such that the Compton length for the Hubble mass is given by the dS curvature.
{}~
\\{}~
\par {\large{\bf Acknowledgments}}{} ~\\{}~\par

This work was supported (F.T.) by Edital Universal CNPq, Proc. 472903/2008-0.
J.P.G. is grateful to CBPF, where this work was completed, for hospitality.

 \end{document}